\documentclass{vgtc}
\ifpdf
  \pdfoutput=1\relax                   
  \usepackage{graphicx}                
  \DeclareGraphicsExtensions{.pdf,.png,.jpg,.jpeg} 
\else
  \usepackage{graphicx}                
  \DeclareGraphicsExtensions{.eps}     
\fi%

\graphicspath{{figures/}{pictures/}{images/}{./}} 

\usepackage{microtype}                 
\PassOptionsToPackage{warn}{textcomp}  
\usepackage{ulem}
\usepackage{comment}
\usepackage[explicit]{titlesec}
\usepackage{adjustbox}
\usepackage{textcomp}                  
\usepackage{mathptmx}                  
\usepackage{times}                     
\usepackage{cite}                      
\usepackage{tabu}                      
\usepackage{booktabs}                  

\usepackage{wrapfig}
\usepackage{subfig}
\usepackage{tabularx}
\usepackage{graphicx}
\usepackage{xcolor}
\usepackage{enumitem}
\usepackage{comment}

\usepackage{flushend}

\newcommand{\new}[1]{\textcolor{black}{#1}}

\onlineid{0}

\vgtccategory{Research}

\vgtcinsertpkg

\title{Data Feel: Exploring Visual Effects in Video Games
to Support Sensemaking Tasks}

\author{Hongwei Zhou\thanks{e-mail: hzhou55@ucsc.edu}
\qquad  Angus G. Forbes\thanks{e-mail: angus@ucsc.edu}
\\{\scriptsize Department of Computational Media}%
\\{\scriptsize University of California, Santa Cruz}%
}

\abstract{
This paper explores the use of visual effects common in video games that support a range of tasks that are similar in many ways to analysis tasks supported in visual analytics tools. While some visual effects are meant to increase engagement or to support a game's overall visual design, we find that in many games visual effects are used throughout gameplay in order to assist a player in reasoning about the game world. In this work, we survey popular games across a range of categories (from casual games to ``Triple A'' games), focusing specifically on visual effects that support a player's sensemaking within the game world. Based on our analysis of these games, we identify a range of tasks that could benefit from the use of ``data feel,'' and advocate for the continued investigation of visual effects and their application in data visualization software tools.
} 

\CCScatlist{
  \CCScat{}{Human-centered computing}{Visualization}{Visualization techniques}{}.
}

\titleformat{name=\paragraph,numberless}[runin]
  {\normalfont\normalsize\bfseries}{}{1pt}{\uline{#1.}}

\begin{document}

\firstsection{Introduction}

\maketitle

The overarching trend in data visualization is to design software tools using a minimalist aesthetic so as not to introduce visual clutter that could impede a data analysis. For example, Tufte famously advocates for a low ``data-to-ink ratio'' in which ``nothing can be erased without losing information''~\cite{tufte2001visual}. While the use of what has come to be called ``chartjunk''~\cite{bateman2010useful,few2011chartjunk} has shown to improve recall and engagement in some cases~\cite{borkin2013makes,borkin2015beyond,gough2014art}, the prevailing wisdom regarding representing information for analysis purposes adheres to these minimalist guidelines, as extraneous visual elements can interfere with efficient visual processing, both in static and interactive visual representations~\cite{ware2019information}. Munzner presents a series of ``rules of thumb,'' summarizing empirical evidence related to the perception of various data representations and cautioning against the use of effects that ``impose significant cognitive load on the viewer''~\cite{munzner2014visualization}. In particular, she warns against the cognitive overload that can occur when animations simultaneously change across different views, as users are blind to regions that they are not currently focusing on, which may cause them to rely on memory rather than their eyes. At the same time, Munzner underscores the importance of visual feedback and responsive interactions, ideally occurring ``immediately'', or in less than 1 second. To reduce interaction costs, she advocates for designing visualization software that explicitly makes use of effective visual encodings to draw attention to automatically detected features, while at the same time acknowledging that the complexity of many analysis tasks requires a human in the loop. 

On the other hand, many video games take advantage of a much richer palette of visual elements, seemingly without hindering a player's ability to carry out various game tasks. Animated visual effects (or ``vfx'') are an integral component of most video games, and their use determines how a player interprets and responds to the various elements within the game. That is, beyond aesthetics, and in addition to enhancing player engagement, the use of visual effects is an essential component of game design that facilitates sensemaking~\cite{pirolli2005sensemaking}, \textit{despite} the visual complexity that it introduces. A popular game design framework called MDA (which stands for ``mechanics, dynamics, and aesthetics'') articulates the interrelationships between the different game elements, explaining that aesthetic choices necessarily have an impact on data representation, algorithms, and player behavior~\cite{hunicke2004mda}. Swink examines what he calls ``polish'', which he describes as the effects that are needed for a player to create a “detailed, expansive mental model” of the game world, explaining the importance of incorporating congruent polish to create an effective interactive experience~\cite{swink2008game}. Specifically, he discusses the significance of visual effects as ``temporary indicators of interaction and movement'', such as particles, trails, and sparks, which enable a player to infer ``a universe of possible interactions'' from a small set of observations. 

In this paper, we explore visual effects in games in terms of their ability to aid a player in reasoning about a complex virtual environment. We specifically focus on in situ vfx that are part of the main action of the game itself, rather than cut scenes or transitions. Rather than comprehensively surveying a wide range of games, we instead choose to focus more in depth on a few popular recent games across four different categories (puzzle, platformer, action, and strategy). Users, once familiar with the gameplay, become adept at processing even very busy animations and visually saturated games scenes, and are able to quickly analyze information in order to make both split second decisions and longer term strategic decisions. Moreover, these visual effects are also used to impart information about how to interact with the game environment, signaling meaningful events and articulating the underlying game mechanics.

Section 2 frames our contribution in terms of previous research in visualization related to animation and visual effects. Section 3 investigates a range of games across multiple genres and presents categories of uses of visual effects in games. Section 4 discusses opportunities for visualization designers to incorporate ``data feel'' into data visualization projects.


\section{Related Works}


\subsection{Animation \& Motion in Data Visualization}

Hubert \& Healy~\cite{huber2005visualizing} find that motion coding is independent from color and shape coding and that more subtle motions are less distracting to users, while still easily perceived. Through a series of experiments that analyze various aspects of motion--- velocity, direction, and on-off blinking--- they demonstrate that each of these properties are effective for encoding multiple data values in a prototype astrophysics simulation, provided they are above particular thresholds of perceptibility. Bartram et al.~\cite{bartram2003moticons} show that brief, simple motions are perceptually efficient ways to distinguish objects in a crowded display, and explore how motion maps to perceptual cues. Ware \& Bobrow~\cite{ware2006motion} introduce a technique termed ``motion highlighting'' to explore the use of motion in node-link diagrams, finding that translating or scaling nodes can be more useful for supporting rapid interactive queries on node-link diagrams than static highlighting methods. Inspired in part by character animations and visual effects in video games, Lockyer \& Bartram~\cite{lockyer2012affective} examine how ambient visual cues can evoke a range of emotional responses in visualizations, and find that variations in path curvature, speed, and texture layout within motion textures can influence affective impressions. Chalbi et al.~\cite{chalbi2019common} explore the Gestalt law of common fate applied to the position, size, and luminance of data objects. A survey paper by Chevalier et al.~\cite{chevalier2016animations} updates the influential taxonomy introduced by Baecker \& Small~\cite{baecker1990animation}, identifying  23  ``roles'' of animation commonly encountered in visual interfaces, such as ``revealing data relationships'' and ``illustrating an algorithm.''

Researchers have studied the use of motion within interactive data visualizations to highlight particular data types or clusters~\cite{lu2020enhancing,etemadpour2017density,veras2019saliency,smc2013}. Chevalier et al.~\cite{chevalier2014not} look specifically at ``staggered'' animations, finding inconclusive results about their potential benefits (such as reducing visual occlusion). Robertson et al.~\cite{robertson2008effectiveness} note that while users found a Gapminder style data animation to be ``enjoyable and exciting,'' it was ``the least effective form for analysis'' when compared to two static visualizations. Nonetheless, a range of projects effectively feature animation as a primary component for presenting streaming data. For example, Huron et al.~\cite{huron2013visual} introduce a design metaphor that presents moving pieces of data falling from the top of the screen into a bar chart, pie chart, or bubble chart representation to illustrate simultaneously both the flow of information over time and an aggregate snapshot of the current state of the data. Other projects that use animation in creative ways to represent patterns in streaming data include visualizations of the stock market using information flocking~\cite{moere2004time}, animated cartograms of urban traffic~\cite{nagel2017shanghai,cruz2015contiguous}, and creative representations of evolving cultural data sets~\cite{LegradyLeonardo2017,ForbesBehaviorism2010}.

\subsection{Game Visualization} \label{subsec:GameVis}
	
Medler \& Magerko~\cite{medler2011analytics} discuss the use of ``playful infovis'' within games to help players make sense of various game statistics to facilitate tracking competition between players and teams. Zammitto~\cite{zammitto2008visualization} explores game interface design in games where players navigate complex 3D worlds, such as the use of heads-up displays, minimaps, and other 2D overlays for wayfinding and resource management tasks. 

Game analytics practitioners make use of various visualization techniques to analyze player behavior, to tune game strategies, and to balance resource allocation systems. For example, Drachen \& Schubert~\cite{drachen2013spatial} investigate the use of heat maps and trajectory visualizations for analyzing spatiotemporal patterns in video games. More sophisticated visual analytics approaches are introduced by Nguyen et al.~\cite{nguyen2015glyph}, which presents a network visualization tool called \textit{Glyph} to analyze problem solving strategies in puzzle games. Ahmad et al.~\cite{ahmad2019modeling} utilizes visualization tools as part of their Interactive Behavior Analytics methodology to model individual and team strategies, and Klein et al.~\cite{kleinman2020and} makes use of a visualization tool called \textit{StratMapper} to reason about sequences of users’ game behaviors.

\subsection{Visual Effects Analysis} 
Although visual effects are a crucial component of video games, exploring the sensemaking aspects of these effects is not a main focus of research within the technical games research community. Rather, analyses of game effects tend to focus on, for instance, on understanding the psychological effects of gameplay~\cite{isbister2016games}, cultural representations in games~\cite{murray2017video}, or a game's underlying operational logics~\cite{wardrip2020pac}. Manovich~\cite{manovich2002language} considers games as a form of new media, extending traditional film analysis approaches to make sense of new types of narrative that emerge when navigating different types of ``computer space'', such as the 3D environments presented in the video games \textit{Myst} and \textit{Doom}. 

Kucic~\cite{gray_juice} describes strategies for successful prototyping emphasizes the importance of utilizing visual effects to add  ``juice'' into a video game so that it ``feels alive and responds to everything you do.'' These effects serve to provide feedback to the player ``by constantly letting them know on a per-interaction basis how they are doing.'' An instructive Game Developers Conference presentation by Jonasson \& Purho~\cite{juiceGDC_Jonasson_Purho} demonstrates a practical example of adding simple but powerful visual effects to a customized version of the early Atari game \textit{Breakout}. Without altering the underlying mechanics of the game, they successively add responsive effects, such as a particle system that sparks when the paddle is hit or animations that play as a brick is disintegrated. Although each additional effect increases the overall visual clutter, this does not at all impede the user from navigating the game. In fact, because each of the visual effects reinforces the game mechanics, they increase a player’s engagement and improves the player's focus. A study by Hicks et al.~\cite{hicks2019juicy} similarly finds that visual embellishments improve player experience of video games.

Swink's influential text \textit{Game Feel} explicitly breaks down various aspects of game design into a series of guidelines--- what he terms input, response, context, polish, metaphor, and rules metrics--- which, he argues, must be simultaneously developed in order to give a game ``juiciness'' and provide an enjoyable, coherent, and immersive experience for a player. Specifically, Swink analyzes games with visual effects of varying degrees of complexity, such as \textit{Asteroids}, \textit{Super Mario Brothers}, and \textit{Gears of War}, and develops a set of high-level principles to promote the design of ``virtual sensation'' within video games. More recent work by Pichlmair \& Johansen examines the game feel of contemporary games related specifically to moment-to-moment interactivity and microinteractions, focusing on examples of visual effects used to highlight particular character movements (e.g., ``invincibility frames''), game events (e.g., ``screen shake''), time manipulation (e.g. ``bullet time''), and object persistence (e.g. ``particle trails'') \cite{pichlmair2021designing}.  

Hubert-Wallander et al.~\cite{hubert2011stretching} summarize research into the positive effects that playing fast-paced video games has on increased visual attention, enabling players to more effectively identify and process endogenous and exogenous data cues. 
Milam et al.~\cite{milam2011effect} analyze camera movement and object motions in video games, exploring how repeating, harmonic, or rhythmic motions can mitigate visual complexity. Milam et al.~\cite{milam2013visual} further explore how game designers aggregate low-level motion features, including speed, size, and density, into meaningful visual effects that help users differentiate game elements and ignore distractions.

\section{Uses of Visual Effects in Video Games} \label{gamevisualeffect}

Here we explore different uses of visual effects that can assist sensemaking tasks based on our survey of video games across multiple genres. These uses are not exclusive to each other. Rather, they are distinguished based on the different intentions a designer might have when utilizing visual effects. While we focus on a few examples of games with pronounced visual effects in particular genres, we believe the uses identified here are shared in many other games, and in the attached appendix after the references we provide a table with additional examples, along with associated images and video links. 

\subsection{Drawing Attention}

%
%

Given the number of events that can occur simultaneously during gameplay, a common use of visual effects in games is simply to direct the user to pay attention to important game elements. This relates to a larger categories of visual techniques that try to establish relations between visual elements, such as camera framing or dynamic lighting~\cite{el2009dynamic}. We find that the uses of \textit{drawing attention} specifically serves to establish a demarcation between foreground and background elements, providing important clues to the player about which visual elements are relevant for gameplay.


\textit{Idle animations} typically describe the animations of the playable character when the player stops moving~\cite{vacheron_2017}. We also use this term to refer to the subtle cuing effects attached to any game entity that is not currently being interacted with. These effects can automatically activate periodically, or in response to a pause in user activity, offering a subtle hint about additional interaction options available in the game. For example, idle animations are used in \textit{Sokobond}, a minimalist puzzle game in which the player manipulates chemical elements within a grid of cells to create composite substances. The corner points of the grid cells include specific interaction mechanics that let the user break or strengthen bonds between chemical elements. When a relevant corner point is ignored, the corner points begin to pulsate rhythmically, evoking a ``breathing'' effect that contrasts with the other static game elements, such as borders and grids. That is, this animation serves as a non-intrusive reminder to the players that these corner points are not part of the background, but rather are meant to advance the gameplay. 

Similarly, in the 2D platforming game \textit{Celeste}, the player controls the character Madeline to get past levels using different abilities, such as jumping and dashing. A range of game elements utilize idle animations through rhythmic up-and-down bouncing motions. These include strawberries (which are key collectibles in the game), green diamonds (which recharge the player's ability to dash quickly), as well as speech bubble icons (which indicate that a character can be spoken to). These effects draw the user's attention to particular game elements, inviting player interaction, and moreover they also serve a clear visual separator of foreground and background elements, indicating which elements are important for gameplay versus which elements can be ignored.

\begin{figure}[tb]
\begin{center}
\includegraphics[width=\columnwidth]{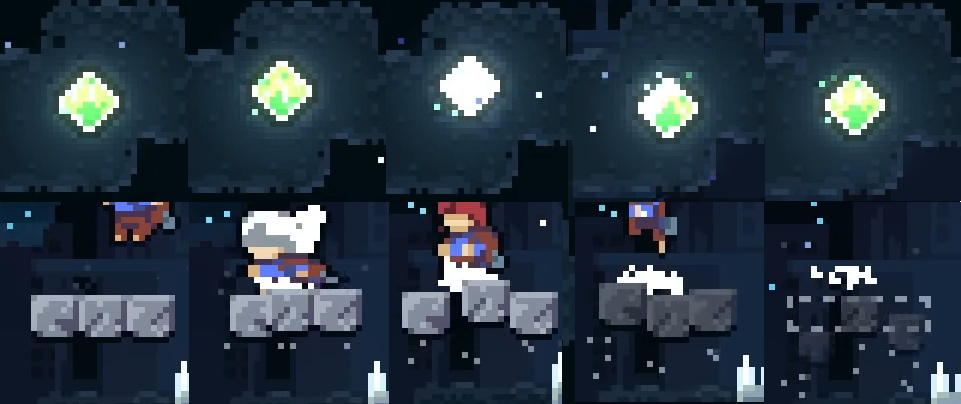}
\vspace{-6mm}
\caption{\textit{Celeste}'s green diamond idle animation (top): It hovers up and down, along with white highlight and green particles. (bottom): The animation of each stone shaking, signaling stone platform collapse .}
\label{fig:CelesteDS}
\vspace{-9mm}
\end{center}
\end{figure}

In addition to animation, many other visual properties, such as color and brightness, serve as key game design considerations to help players understand the game narrative. In \textit{Celeste}, the first level of the game takes place in a city ruin during winter. The color palette of the background is muted and cold, while gameplay-related entities, such as strawberries and green diamonds, stand out due to their bright, warm colors, along with glowing animated effects, as shown in Fig.~\ref{fig:CelesteDS}. The green diamonds are represented using particle effects, which emphatically draw a player's attention, and are also inherently visually pleasing, positively reinforcing the player's desire to interact with them. In contrast to the first level, later in the game the background environment becomes more warm, lively, and colorful. This progression from a muted to saturated color palette helps players first situate themselves in the game world and become familiar with the visual language of the game. Once a basic mastery of game interactions is established and a clear delineation between foreground and background elements is made, more visual variety can be introduced without confusing the player.

\subsection{Representing Mechanics}

The term ``mechanics'' is used to describe the rules that govern the actions and goals of a game. Zubek~\cite{zubek_2020} defines game mechanics as comprised of three basic components--- \textit{entities}, \textit{actions} performed by or on entities, and \textit{rules} that describe the conditions and outcomes of those actions--- that describe the nouns, verbs, and grammar of the game, respectively. Hunicke et al.~\cite{hunicke2004mda} further contextualize mechanics as existing ``at the level of data representation and algorithms.'' In order to play a game successfully, the user has to understand the range of possibilities available within the game, which are dependent on the underlying game mechanics. While players may be familiar with aspects of mechanics associated with particular game genres, games that efficiently introduce new mechanics successfully help the user become familiar with them through a range of visual cues. Indeed, part of the enjoyment of playing a video game is gaining a mastery of these mechanics and taking advantage of them to explore the various possibilities that they enable.

\begin{figure}[tb]
\begin{center}
\includegraphics[width=\columnwidth]{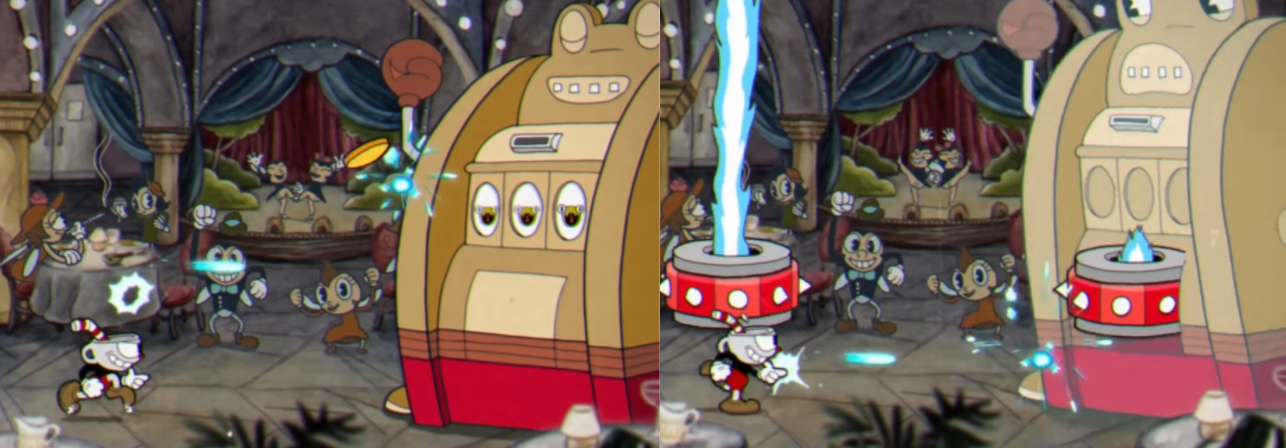}
\vspace{-6mm}
\caption{In \textit{Cuphead}, blue particles (found beneath the handle on the left subfigure, next to the red base on the right subfigure) indicate shot hit. The white overlay on the machine on the right subfigure indicates damage done.}
\label{fig:CupheadHighlight}
\vspace{-9mm}
\end{center}
\end{figure}

A major component of game design involves guiding the player to ``think'' using the logic of the game's mechanics. Visual effects can provide a consistent visual language that helps to establish the rules of gameplay. For instance, the game \textit{Cuphead} is visually rich, with a unique hand-drawn aesthetic that evokes the style of early cartoons, and uses a wide range of visual effects to ``explain'' various aspects of the game logic. During a battle, when a player shoots, a blue explosion effect is used to indicate that the shot hits the enemy. However, the shot landing does not necessarily lead to the enemy taking damage (since some enemies can become temporarily invincible or have protected spots). Therefore, another distinct visual effect where the enemy flashes white is used to communicate damage done. As shown in Fig.~\ref{fig:CupheadHighlight}, the two visual effects--- the explosion of blue particles when a shot lands, and the white flashing to indicate damage--- are essential for providing user feedback about their actions during the battle. Unlike ``background'' visual effects that are used to articulate overall game aesthetics, each game mechanic effect tends to indicate a single and specific meaning, continually reinforcing the logic of the game.



Game mechanic effects are used extensively in strategy games. In the popular \textit{Total War} series, an entire army is represented by a single character on the game's ``strategy map.'' Attrition mechanics are invoked when an army enters into a disadvantaged environment, such as harsh winter snow or lands affected by ``vampiric corruption.'' This is typically represented by distinctively negative imagery and visual effects, such as animations that display red sparks and skulls or animations of wounded people stumbling and falling over. During battles scenes in \textit{Total War}, each unit has a single number representing its ``morale.'' When the morale falls below a certain point, the entire unit stops fighting and starts to retreat from the battlefield. In \textit{Total War: Three Kingdoms}, this is represented by a flashing flag icon on the unit, providing a warning to the player. These effects clearly distinguish these units from those still engaged in battle, whose icons are represented with solid, static colors. This visual language articulates the logic of the battle, helping the player to quickly dissect the battle scene in order update their strategy, if necessary.


\begin{figure}[tb]
\begin{center}
\vspace{-4mm}
\includegraphics[width=\columnwidth]{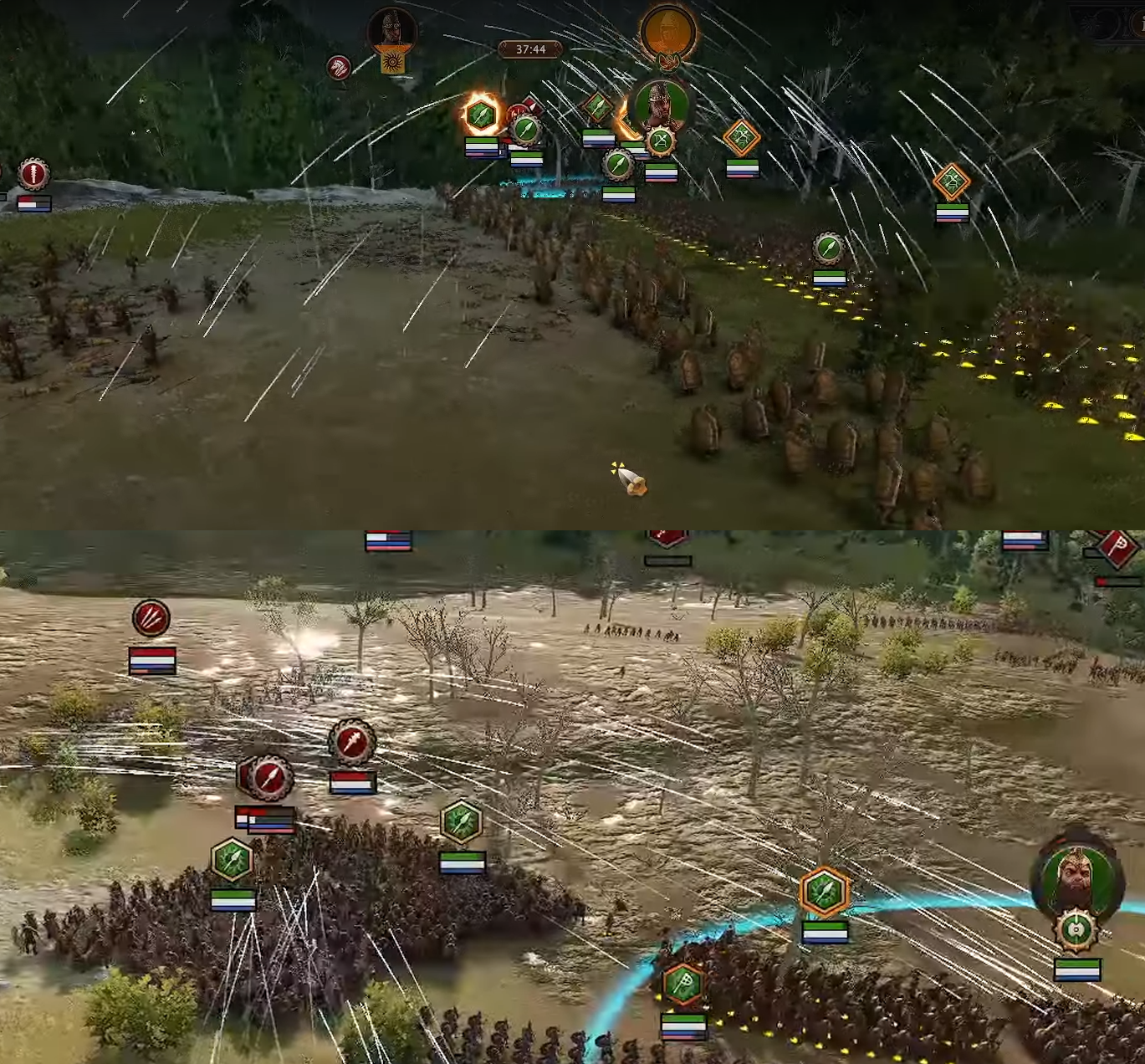}
\vspace{-6mm}
\caption{Two different archer fire modes in \textit{Total War: Troy}. The top screenshot shows arc fire to avoid hitting friendly troops at the front line. The bottom screenshot shows direct fire when the archers have a direct line of sight to their targets.}
\label{fig:TotalWarArrow}
\vspace{-8mm}
\end{center}
\end{figure}

In addition to representing a specific unit of meaning, the use of a visual effect can also impart more nuanced information. This is exemplified by certain attributes of the visual effects, such as length, duration, and density being mapped to particular attributes within the game systems, such as range of the dash mechanic in \textit{Celeste}, the range of explosions in \textit{Hades}, or effectiveness of archer firing in the \textit{Total War} series. This is akin to visualization studies that explore how particular attributes of motion, such as amplitude, frequency and phase, can be used to encode data~\cite{ware2006motion}.

\begin{wrapfigure}{l}{0.4\columnwidth}
\begin{center}
\vspace{-4mm}
\includegraphics[width=0.4\columnwidth]{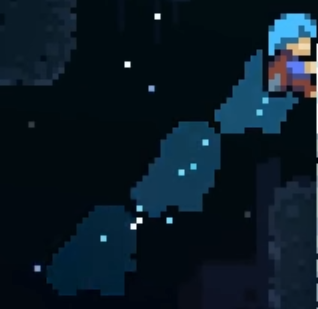}
\vspace{-6mm}
\caption{The player silhouettes as a trail after dashing in \textit{Celeste}, providing insight into properties of the game's dash mechanic.}
\label{fig:CelesteTrail}
\vspace{-5mm}
\end{center}
\end{wrapfigure}

In \textit{Celeste}, the player can quickly ``dash'' in any of eight different directions to manipulate their position, even while in mid-air. This action is accompanied by a visual animation of a trail, which takes the form of several player silhouettes, as shown in Fig.~\ref{fig:CelesteTrail}. Without these effects, it would be difficult to track the range and travel trajectory of the dash simply because of how quickly the dash happens. Similarly, a visual effect is used in \textit{Hades} when a magical explosion occurs, outlining a colored circular boundary to indicate which area is directly impacted by the explosion, as shown in Fig.~\ref{fig:HadesCuphead}(right). These correlations between the player's abilities and their visual representation enable the player to assess and master the available actions in the game system.

During the battles in the \textit{Total War} series, artificial trailing effects are added to long range trajectories, such as arrows and javelins. This effect is a particularly helpful form of visual communication, as the Total War games feature sophisticated simulations of ranged weaponry. As shown in Fig.~\ref{fig:TotalWarArrow}, the archers have two modes of shooting: either direct shots if enemies are in direct lines of sight (bottom), or arched shots if no direct sighting (top), which typically takes more travel time and has less accuracy. The density of the arrow trail also indicates different qualities of archer units, such as their firing rate. When the player orders an attack order but sees very few arrow trails, they might suspect that the archers are not in a good position to fire, due to the terrain or line of sight being blocked. The most recent entry \textit{Total War: Warhammer III} also introduced lazy health bar, which reacts to ``incoming damage by highlighting the new value and draining to it over time, helping highlight sudden large bursts of damage''~\cite{total_war_2021}. Similar to the use of the trail effect in \textit{Celeste}, changes in state over time are communicated through visual effects.

\subsection{Signaling Events}

\begin{figure}
\begin{center}
\vspace{-4mm}
\includegraphics[width=\columnwidth]{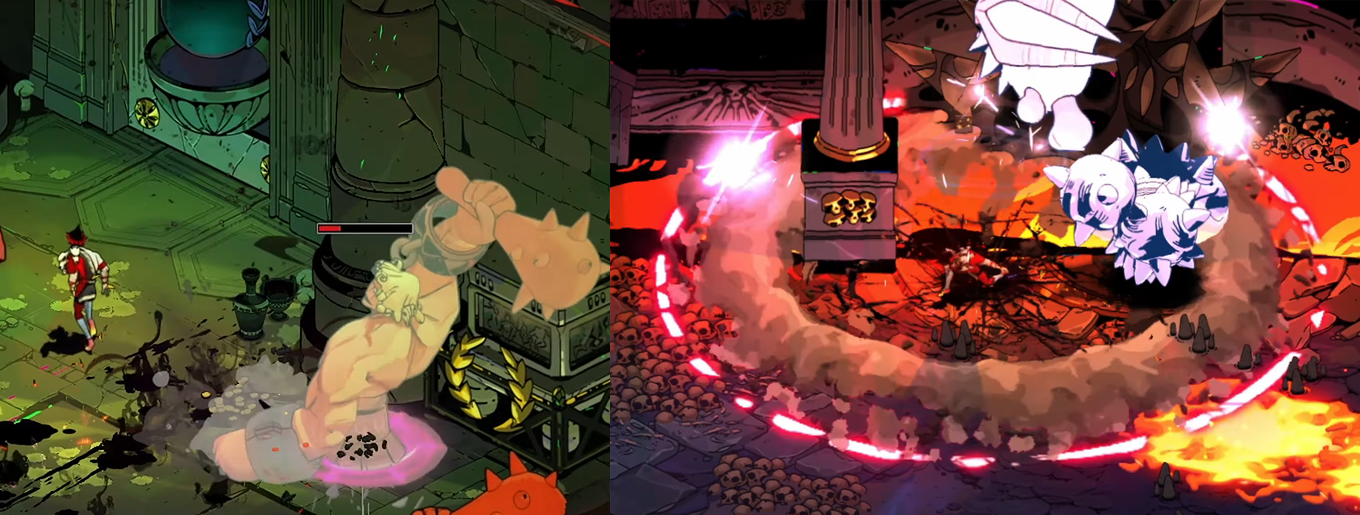}
\vspace{-6mm}
\caption{(left): In \textit{Hades}, the enemy holds a pose and is colored with a white overlay to signal that an attack is imminent. (right): The magic explosions in \textit{Hades} have easily discernible outlines to clearly indicate their area of impact.}
\label{fig:HadesCuphead}
\vspace{-8mm}
\end{center}
\end{figure}

Visual effects can also signal that important gameplay events are either about to take place or have just happened. In the context of video games, events refer to specific occurrences that change the state of the game, typically actions taken by an entity (e.g, attacking an enemy) or interaction between two entities (e.g., the player took damage after falling on spikes). 

One of the most common uses of a signaling effect is to indicate a meaningful upcoming event, preparing the player to transition to a different modality of gameplay. In action games, this is also called ``telegraphing,`` where the enemy characters have a recognizable animation or pose before the attack actually hits, such as crouching into a fighting stance or brandishing their weapons. In \textit{Hades}, when an enemy is about to attack they become highlighted in white (Fig.~\ref{fig:HadesCuphead}, left). In \textit{Celeste}, there are stone platforms where the player can only step on it once before they collapse. The stone floor plays a shaking animation before it disappears as a warning to the player (Fig.~\ref{fig:CelesteDS}, bottom).

This type of visual effect can also signal events that already happened, mostly commonly used to communicate entity interactions or to provide visual feedback to player actions. For example, in \textit{Sokobond}, when a corner point is used by the player--- such as when a bond is cut or strengthened--- the corner point expands suddenly, then quickly returns its original size. This ``jumping'' animation not only draws the player attention to the element, but also establishes a connection between the corner point behavior and the chemical substance in the cell. Without this motion, one can imagine that bonds are cut without the player realizing it happening or why it happened. Similarly, in \textit{Cuphead}, the player is able to ``parry'' certain entities in the game to collect ``charges'' that can be used to carry out powerful attacks needed to defeat certain enemies. These parry events are treated as significant moments in the gameplay. In addition to visualizing a purple glow where the parry occurs, the entire game freezes for a moment to emphasize the importance of the event. Visual effects that signal events add duration to an otherwise instant occurrence within the game, and these temporal fluctuations are thus part of the visual language of the game system. 

\section{Applying Visual Effects to Data Visualizations} \label{sec:definingVisualEffect}


Video games and data visualizations of course have entirely different purposes. We argue that an understanding of the different types of game effects could be leveraged by visualization designers to create more engaging interactive visualization to support various sensemaking tasks. For example, the use of idle animations, used widely in games, are very effective at reminding a player of what interface options are available and what game elements are useful. Rather than distracting the player from completing particular game tasks, they support them. That is, they strike an appropriate balance that triggers the player's awareness without being distracting. The various ``drawing attention''  mechanisms found in minimalist puzzle games, such as \textit{Sokobond}, \textit{Mini Metro}, or \textit{Wordscapes}, seem particularly amenable to visual analytics dashboards. Referencing Swink's \textit{Game Feel}, these subtle reminders have the potential to give the impression that the entire visualization space is ``a universe of possible interactions,'' directing the user to various functionality available to support particular analysis tasks. Moreover, if data visualization tools incorporated these types of visual effects, this would make it clear which elements could be considered foreground, requiring attention, versus those that were in the background. From our analysis of video games, it seems clear that the reason that players can navigate busy visual scene in games is precisely because they successfully define a visual language that functions to separate foreground from background elements, often in very sophisticated ways, helping players to reason strategically about the game elements, and changing as the various contexts (dwindling resources, presence of enemies, etc.) update over the course of the game. 

Visual effects can be used to represent game mechanics, whereby particular rules or procedures of the game systems are communicated. The player is (ideally) able to effortlessly pick up the visual language via these effects and generate reasonable hypotheses about how the internal system functions. In the context of data visualization, it could be useful for visualization designers to think about how important parameters or processes can be obscured from the user, and how these could be communicated through the use of visual effects. Visualization designers can identify and expose hidden intricacies that can assist users to better make sense of the visualization tool and the data it presents. While data visualization tools tend not to take advantage of what we are calling ``data feel,'' some interesting projects do include useful visual effects. For example, Suh et al.~\cite{suh2019persistent} describe an interface that presents a sophisticated physics simulation of force-directed graphs. Through this interface, the user can customize the attracting and repelling forces of nodes to visualize different clustering results. In this case, the specific mechanics of force acting between nodes are hidden, only inferred by their movements and final clustering results. Additional visual effects, such as color on the edges that communicate the acting of forces, can clarify underlying systemic processes without being distracting, supporting rather than interfering with the data analysis process.

Visual effects used for event signaling in games also could be usefully adapted to data visualization contexts. Specifically, and perhaps counterintuitively, visual effects can provide meaningful feedback to user actions especially when the interface is busy. For instance, many video games maintain an awareness of the player's current situation, and signal to the user, for example, when they are in imminent danger, or when an achievement is about to be unlocked. In a similar manner, we can conceive of a visualization tool that is aware of the potential for introducing bias into a data analysis session, or that is able to sense that a particular modeling choice would lead to statistical significance.

More generally, designers can utilize visual effects to better organize the interface without obscuring information. Many visualization tools focus on offering comprehensive views and menus of interactive options that can unintentionally overwhelm the user. Based on our survey in Sec.~\ref{gamevisualeffect}, visual effects can serve to not only prioritize different visual elements on screen, but also provide visual feedback to user interactions and to systemic events. In both of these cases, visual effects promote sensemaking and assist with selective attention, including the selection of on-screen elements (foreground vs. background) and the selection of relevant elements over time (mechanics and events). Game effects can enable designers to embed additional interactive functionality, thus allowing more ``spaces'' in which to present information and to create new kinds of user experiences.

\subsection{Planning vs. Execution}

Many of the visual effects investigated in \autoref{gamevisualeffect} demand a player reaction within a short span of time, especially for effects representing signaling events, such as those used for telegraphing imminent enemy attack in \textit{Hades}. Even for effects that are meant to emphasize certain aspects of the game, such as the arrow trails in \textit{Total War}, the intention is to provide relevant feedback so the player can quickly assess the battle situation and make decisions. This is distinct from other player actions where situation assessment is less urgent. Atymiz et al.~\cite{aytemiz2020diagnostic} distinguish between \textit{planning} and \textit{execution} interactions. The games with more demand on execution typically require the player to react to the situation on the fly, such as in action and shooting games, while those with more demand on planning require the player to learn the game system, such as in puzzle games.

With this distinction in mind, we can see that the goals of many visualization tools are mainly focused on planning: the user is mostly responsible for learning the mechanics of the visualization tool, generating and testing out hypotheses without any introduction of time constraints. This suggests that visualization tools with real time components, such as, for example, interfacing with robots remotely or monitoring disasters or outbreaks, are the more salient areas to experiment with visual effects related to execution. Additionally, other visualization tools may also have real time execution demands, such as those that involve intricate modeling or simulation systems.

\subsection{Juice, Oil, and Pleasurable Interaction}

In the game literature we surveyed, ``juice,'' ``polish,'' or ``game feel'' is often used to describe a wide variety of improvements to gameplay, and is not necessarily limited solely to simple visual embellishments. Pichlmair \& Johansen~\cite{pichlmair2021designing} provide a taxonomy of techniques that potentially can be used to improve game feel. In addition to visual effects, such as screen shake and recoil (of guns) to signal particular events, they include adjustments that are concerned with non-visual controls and game mechanics. For example, the \textit{coyote time} effect refers to a small duration where the player's avatar can still turn, walk, and jump after they maneuver off a cliff. Another example is \textit{corner correction}, which refers to automatically adjusting the player's avatar position when walking into a corner to prevent it from getting stuck when the player fails to walk around the corner perfectly. Many of these improvements can be considered ``oil'' (that makes play smoother) instead of ``juice'' (that makes play more engaging), in Swink's terminology. Both juice and oil are important to create a more pleasurable interactive experience. 


On a similar note, Elmqvist et al.~\cite{elmqvist2011fluid} argue for the importance of creating compelling interactions in information visualization tools. They provide a series design guidelines for ``fluid interactions,'' including immediate visual feedback and visceral rewards for successful interaction to reinforce the visualization's conceptual model. The immediate feedback provides users with a sense of embodiment and immersion so as to minimize ``the gulfs of evaluation and execution'' and to help facilitate states of ``flow.''  Taking inspiration from game effects can further this line of inquiry in visualization design.

\section{Conclusion}

In this work, we survey a range of contemporary video games and identify three non-exclusive uses of visual effects beyond aesthetics and entertainment: \textit{drawing attention}, \textit{representing mechanics} and \textit{signaling events}. We explore the similarities and differences of goals between video games and visualization tools and speculate on how visual effects could assist in sensemaking tasks. Overall, we provide an initial investigation of visual effects in games and advocate for further research into the challenges and opportunities in incorporating ``data feel'' into visualization tools by adopting elements from video game design into the design of data visualization interfaces.

\bibliographystyle{abbrv-doi}

\bibliography{bibliography}

\begin{figure*}
\vspace*{-40mm}
\begin{center}
\hspace*{-135mm}
\includegraphics[width=5.3\columnwidth]{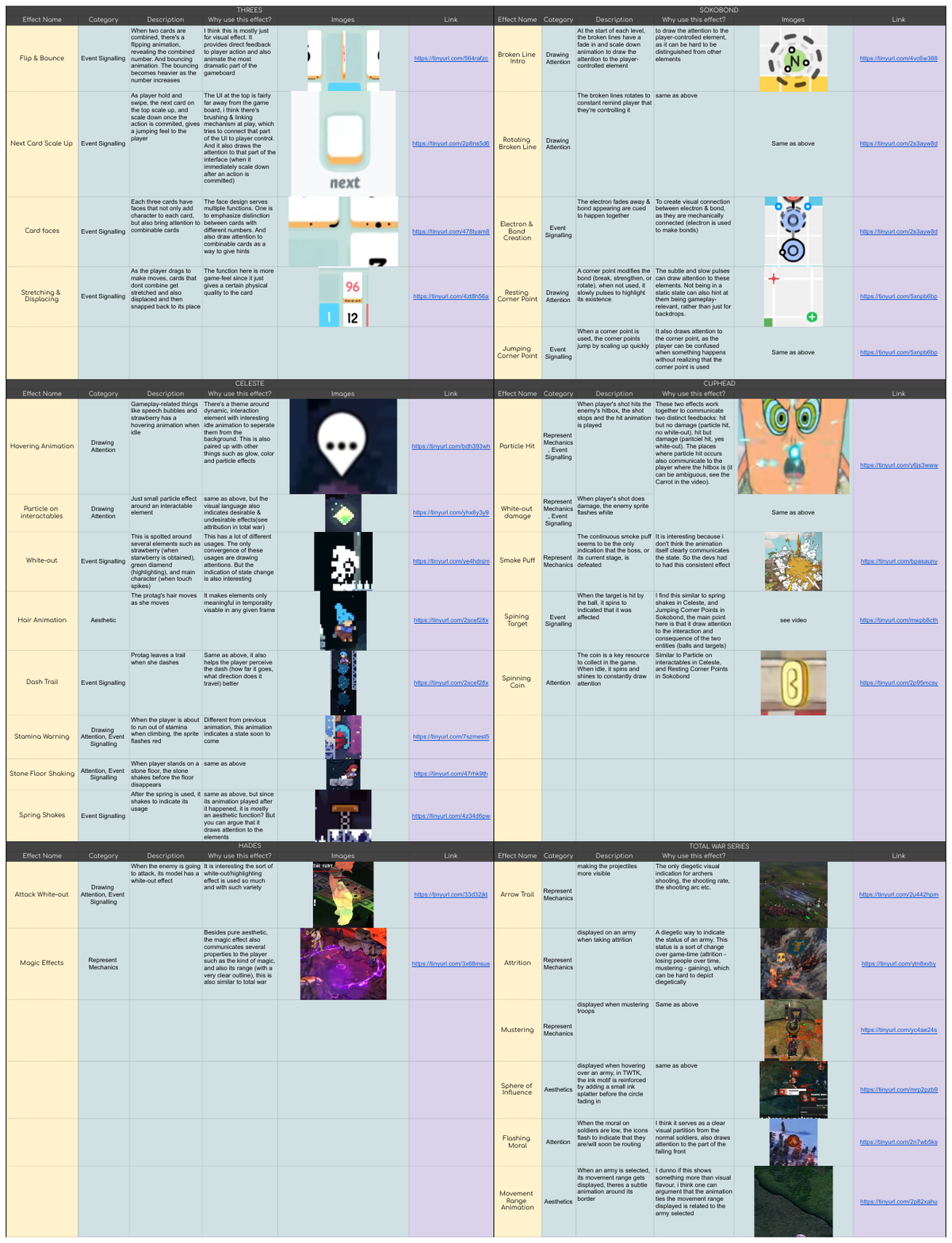}
\label{fig:graphicssurvey}
\end{center}
\end{figure*}

\end{document}